\documentclass[journal=jctc,layout=twocolumn,manuscript=article]{achemso}
\SectionNumbersOn 

\usepackage[T1]{fontenc} 		
\usepackage{amsmath,amssymb}
\DeclareMathOperator{\Tr}{Tr}
\usepackage{bm}
\usepackage{changepage}
\usepackage[utf8x]{inputenc}
\usepackage{nameref,hyperref}
\usepackage{microtype}
\DisableLigatures[f]{encoding = *, family = * }
\usepackage[usenames,dvipsnames,svgnames,table]{xcolor}
\usepackage{array}
\usepackage{url}
\usepackage{dcolumn}
\usepackage{graphicx}
\usepackage{multirow}
\usepackage{lineno}
\usepackage{txfonts}
\usepackage{booktabs}
\usepackage{subcaption}
\usepackage{adjustbox}
\usepackage{wrapfig}
\usepackage{siunitx}

\newcommand{\eq}[1]{Equation~\ref{#1}}

\newcolumntype{d}[1]{D{.}{.}{#1}}

\title{Reaction Coordinates for Conformational Transitions using Linear Discriminant Analysis on Positions}

\author{Subarna Sasmal}
\affiliation{Department of Chemistry and Simons Center for Computational Physical Chemistry, New York University, New York, NY 10003}
\author{Martin McCullagh}
\affiliation{Department of Chemistry, Oklahoma State University, Stillwater, OK 74078}
\email{martin.mccullagh@okstate.edu}
\author{Glen M. Hocky}
\affiliation{Department of Chemistry and Simons Center for Computational Physical Chemistry, New York University, New York, NY 10003}
\email{hockyg@nyu.edu}

\date{\today} 

\makeatletter
\let\l@addto@macro\relax
\makeatother
\usepackage[fontsize=11pt]{scrextend}

\begin{document}

\begin{abstract}
In this work, we demonstrate that Linear Discriminant Analysis (LDA) applied to atomic positions in two different states of a biomolecule produces a good reaction coordinate between those two states. 
Atomic coordinates of a macromolecule are a direct representation of a macromolecular configuration, and yet they have not been used in enhanced sampling studies due to a lack of rotational and translational invariance. 
We resolve this issue using the technique of our prior work, where a molecular configuration is considered a member of an equivalence class in size-and-shape space, which is the set of all configurations that can be translated and rotated to a single point within a reference multivariate Gaussian distribution characterizing a single molecular state.
The reaction coordinates produced by LDA applied to positions are shown to be good reaction coordinates both in terms of characterizing the transition between two states of a system within a long molecular dynamics (MD) simulation, and also ones that allow us to readily produce free energy estimates along that reaction coordinate using enhanced sampling MD techniques. 
 \end{abstract}

\section{Introduction}
A large class of enhanced sampling techniques work by biasing a system to explore along a low dimensional set of collective variables (CVs) \cite{henin2022enhanced}.
These methods allow us, in principle, to use the known bias applied to reconstruct the free energy landscape in that low dimensional space. 
In practice, the choice of the CVs is crucial, with an ideal set of CVs allowing the system to explore all relevant states within available simulation time.\cite{henin2022enhanced} 
Recently, extensive effort has been invested in using a variety of machine learning approaches, from very simple to very sophisticated, to determine optimal coordinates for sampling from molecular dynamics (MD) simulation data (Refs.~\citenum{ma2005automatic,hashemian2013modeling,tiwary2016spectral,chen2018molecular,mendels2018collective,mendels2018folding,wehmeyer2018time,piccini2018metadynamics,sultan2018automated,ribeiro2018reweighted,zhang2019improving,wang2020machine,noe2020machine,sidky2020machine,bonati2020data,karmakar2021collective,tsai2021sgoop,hooft2021discovering,sun2022multitask,rydzewski2022reweighted} provide a representative but not exhaustive sample).

One commonly encountered challenge is to compute the free energy path of a transition between two states along a linear dimension that chemists term a reaction coordinate (RC).
For a macromolecule such as a protein, the two states could be configurations for which we have known structures (e.g. the PDB structure of a protein solved with and without a bound ligand), or processes for which one state is known and the other can be at least qualitatively defined (e.g. folding/unfolding or binding/unbinding). 
If a long MD trajectory containing multiple transitions between these states is available, then reaction coordinates could be trained based on the idea that we want to enhance sampling along the slowest modes in the system \cite{tiwary2016spectral,sultan2018automated,chen2019nonlinear,wang2020machine,noe2020machine,bonati2021deep}.
However, having this data is rare, in which case one can try iterative enhanced sampling and learning reaction coordinates with the goal of maximizing the number of transitions between the two states in a fixed amount of simulation time \cite{tiwary2016spectral,ribeiro2018reweighted,chen2018molecular,sidky2020machine,wang2020machine,mehdi2022accelerating}.  

An alternative approach which has shown some success is to train reaction coordinates based on short simulations within the two states, and use a method that produces a coordinate representing the difference between the two sets of data. 
Linear dimensionality reduction techniques such as Principal Component Analysis (PCA) and Linear Discriminant Analysis (LDA) are the simplest approaches to combining a large set of variables defining a system of interest and producing a small set of CVs that characterize the available data. 
While PCA, which produces coordinates that capture the most variance in the data,  has been used for exploration in enhanced sampling simulations, LDA seems to hold more promise as an RC since it is a supervised approach whose goal is to maximally separate different labeled classes of data (i.e. reactants and products).
We describe LDA in full detail in the next section.
In one study, Mendels et al. \cite{mendels2018collective} produced a modified approach to LDA termed harmonic LDA (HLDA, because the covariance matrices in the two different states of interest are combined by a harmonic average rather than a simple sum), and in that work and subsequent ones,\cite{mendels2018folding,piccini2018metadynamics} combined it with Metadynamics (MetaD) to effectively enhance sampling between two states in several different systems.
Later, a neural network was used to combine features before training LDA vectors to produce the reaction coordinate \cite{bonati2020data}.

In  prior examples of reaction coordinate design for free energy sampling of biomolecules that we are aware of, the input features to the method were internal coordinates, or a function of internal coordinates, for the molecule(s) of interest---for example, distances, angles, and dihedrals. 
Often, these could be CVs based not on atomic positions directly, but on coarse-grained (CG) representations of the biomolecule, such as the distance between the centers of masses (COMs) of two different domains, or the distance between the COM of a ligand and certain atoms in its binding pocket. 
This is not surprising, because these often correspond to our physical intuition about what the biomolecular reaction is.
Moreover, internal coordinates are invariant to translation and rotation of the molecule, and thus bias forces applied to these coordinates do not depend on the position or orientation of the molecule.

Recently, we presented atomic coordinates as an alternative set of features to use in the context of clustering biomolecular data \cite{shapeGMM}. 
Atomic coordinates of a subset of atoms, or of beads corresponding to a CG representation of a molecule, offer an alternative to internal coordinates with the advantage that there is little choice in selecting the features to use.
Using a protein as an example, we need only make the standard choice between $\mathrm{C}_\alpha$ atoms, backbone, all heavy atoms, and so on.
Moreover, only $3N-6$ atomic coordinates essentially describe the state of a biomolecular system with $N$ important atoms (but ignoring contributions of solvent, salt, etc.),  whereas use of internal coordinates often results in an over-determined set of features, such as all $O(N^2)$ pairs of distances.
In Ref.~\citenum{shapeGMM} we developed a procedure for clustering molecular configurations into a Gaussian mixture model (GMM) using atomic positions that overcomes challenges of orientational dependence that prevented their use earlier, as described below.
Because a Gaussian mixture model in positions is a natural way to coarse-grain a free energy landscape,\cite{shapeGMM,tribello2010self,InfleCS,giberti2021global} with locally harmonic bins around metastable states, the resulting clustering is a physically appealing definition of the ``states'' one's molecule can adopt.  

However, our Gaussian mixture model still relies on a very high ($3N-6$) dimensional representation of our molecule. 
Given that the output of our clustering algorithm is a set of states each defined by a multivariate Gaussian distribution, LDA is a natural approach to produce a low dimensional representation of our data with large separation between states.
In this work, we first apply LDA to the folded and unfolded states determined from shapeGMM clustering of a long unbiased MD trajectory of a fast-folding protein, and demonstrate that it produces a physically reasonable ordering of states from folded to unfolded. 
We then show that this coordinate is a ``good'' reaction coordinate because the position of the barrier separating folded and unfolded is very close to the location where the system is equally likely to proceed to folded or unfolded (in terms of a committor function to be defined below).
We implement this position LDA coordinate in the PLUMED sampling library, and demonstrate that biased sampling along this coordinate can accelerate transitions between the folded and unfolded states, and produce a qualitatively similar free energy surface as compared to the unbiased trajectory in 3\% of the simulation time, without any additional tuning of the CV. 
Finally, we train a position LDA coordinate on an achiral helical system where data is only available in the left and right handed states, and show that this coordinate also allows us to readily sample between the two states, even though no information about the transition was provided during training. 

\section{Theory and Methods}
\subsection{Molecules in Size-and-Shape Space}
Consistent with our previous work on structural alignment and clustering,\cite{shapeGMM} we consider structures from a molecular dynamics (MD) simulation to be associated with Gaussian distributions in atomic positions.
Structures are represented by $N$ particles (a subset of atoms) using a vector $\bm{x}$ of dimension $N\times3$
which is a member of an equivalence class,
\begin{equation}
[\bm{x}_i] = \{\bm{x}_i\bm{R}_i + \bm{1}_N\vec{\xi}_i^T : \vec{\xi}_i \in \mathbb{R}^3, \bm{R}_i \in \text{SO}(3)  \},
\end{equation}
where $\vec{\xi}_i$ is a translation in $\mathbb{R}^3$, $\bm{R}_i$ is a rotation $\mathbb{R}^3\rightarrow\mathbb{R}^3$, and $\bm{1}_N$ is the $N\times1$ vector of ones. 
$[\bm{x}_i]$ is a point in size-and-shape space\cite{Dryden1998} which has dimension $3N-6$ and is defined as $S\Sigma_N^3 = \mathbb{R}^{3N}/G$ where $G = \mathbb{R}^3\times \text{SO}(3)$ is the group of all rigid-body transformations for each frame with elements $\bm{g}=( \vec{\xi},\bm{R})$. 

Within the shapeGMM framework, the probability density of particle positions is assumed to be a Gaussian mixture,
\begin{equation}
\label{gmm}
P(\bm{x}_i) = \sum_{j=1}^{K} \phi_j N(\bm{x}_i\bm{g}_{i,j}\mid \bm{\mu}_j, \bm{\Sigma}_j),
\end{equation}
where $N(\bm{x}_i\bm{g}_{i,j}\mid \bm{\mu}_j,\bm{\Sigma}_j)$ is the $j$th normalized, multivariate Gaussian with  mean $\bm{\mu}_j$, covariance matrix $\bm{\Sigma}_j$, and weight $\phi_j$ (the weights are normalized such that $\sum_{j=1}^{K} \phi_j = 1$). 
$\bm{g}_{i,j}$ is the element of $G$ that minimizes the Mahalanobis distance between $\bm{x}_i$ and $\bm{\mu}_j$.  Iterative determination of $\bm{g}_{i,j}$ and $\bm{\mu}_j$ is performed in a Maximum Likelihood procedure.\cite{shapeGMM} 

In the current work, we will only consider LDA coordinates learned using data from only two states.
Additionally, we will only consider ``weighted'' alignment of particle positions, which equates to using a Kronecker product covariance (where $\Sigma_j=\Sigma_N \otimes I_3$, for $\Sigma_N$ the $N\times N$ covariance of particle positions) in defining the Mahalanobis distance between frame and average structure as described in detail in Ref.\citenum{shapeGMM}.

\subsection{Dimensionality Reduction using Linear Discriminant Analysis on Particle Positions}
We propose to use LDA directly on aligned particle positions  as a reaction coordinate.
LDA for two states produces the linear model with the maximal inter-average variance while minimizing intra-cluster variance\cite{bishop2006pattern}.
This is achieved by first computing the within-cluster scatter matrix, 
\begin{equation}
    \label{Sw}
    \bm{S}_w = \sum_{i=1}^K \sum_{j\in{N_i}} (\bm{x}_j - \bm{\mu}_i) (\bm{x}_j - \bm{\mu}_i)^T,
\end{equation}
and the between-cluster scatter matrix, 
\begin{equation}
    \label{Sb}
    \bm{S}_b = \sum_{i=1}^{K} (\bm{\mu}_i - \bm{\mu}) (\bm{\mu}_i - \bm{\mu})^T,
\end{equation}
where $\bm{\mu}_i$ is the average structure of cluster $i$ and $\bm{\mu}$ is the global average.  The simultaneous minimization of within-cluster scatter and maximization of between cluster scatter can be achieved by finding the transformation $G$ that maximizes the quantity
\begin{equation}
    \Tr\left( (G^T\bm{S}_wG)^{-1}G^T\bm{S}_bG\right).
\end{equation}
This maximization can be achieved through an eigenvalue/eigenvector decomposition but such a procedure is only applicable when $\bm{S}_w$ is non-singular. The LDA method was reformulated in terms of the generalized singular value decomposition (SVD)\cite{LDA_SVD} extending the applicability of the method to singular $\bm{S}_w$ matrices such as those encountered when using particle positions. 

In addition to employing the SVD solution of to the LDA approach, care must be taken in how particle positions are aligned when performing LDA.  This is evident when one considers the scatter matrices in \eq{Sw} and \eq{Sb}.  
The values and null spaces of these scatter matrices will depend on the specific alignment procedure chosen.
There are three obvious choices for structural alignment prior to LDA: (1) alignment of each frame to its respective cluster mean/covariance, (2) alignment to one cluster or another, and (3) alignment to a global average.
The first choice will lead to scatter matrices with different null spaces for each cluster making their addition in \eq{Sw} unsatisfactory.  Alignment options (2) and (3) will both yield consistent null spaces making the choice of one over the other not immediately obvious.  While there may be reasons to selection option (2), we have chosen to move forward with global alignments for the systems studied here. 

The result of an LDA procedure will be a set of $K-1$ vectors $\{\bm{v}\}_{K-1}$ of coefficients that best separate the data.  
These vectors are similar in nature to the eigenvectors from a principle component analysis, a procedure more familiar to the bio-simulation field. 

\subsection{Biasing a linear combination of positions}
The value of the LDA coordinate after this procedure is a dot product of the vector $\bm{v}$ with the atomic coordinates $\bm{x}-\bm{\mu}$. 
When computing this value on the fly within an MD simulation, we need to consider the value of $\bm{[x(t)]}$, the equivalence class of the position at time $t$, translated and rotated to a reference $\bm{\{\mu,\Sigma\}}$. 

Therefore, to compute the value of the LDA coordinate $l$, we first translate $\bm{x(t)}$ by $\vec{\xi}(t)=\frac{1}{N} \sum_{i=1}^N \vec{x}_i(t) - \frac{1}{N} \sum_{i=1}^N \vec{\mu}_i(t)$, the difference in the geometric mean of the current frame and that of the reference configuration. 
Then, we compute $\bm{R}(t)$, the rotation matrix which minimizes the Mahalanobis difference between $\bm{x(t)}-\vec{\xi}$ and $\bm{\mu}$, for a given $\bm{\Sigma}$, as described in Ref.~\citenum{shapeGMM}. 
Finally, we compute
\begin{equation}
    l(\bm{x}) = \bm{v} \cdot \big(\bm{R}\cdot(\bm{x}(t)-\vec{\xi}(t)) - \bm{\mu} \big)
\end{equation}
By definition, $l(\bm{\mu})=0$.

To apply bias forces to this coordinate, we must be able to compute $\nabla l(\bm{x}$(t)).
Because of the inclusion of the optimal rotation process by SVD, it is non-trivial to compute this analytically, and we instead compute derivatives numerically. 

\subsection{Enhanced sampling with OPES-MetaD}
Enhanced sampling simulations on LDA coordinates were performed using Well-tempered Metadynamics (WT-MetaD), and On the Fly Probability Enhanced Sampling-Metadynamics (OPES-MetaD) as implemented in \texttt{PLUMED} \cite{tribello2014plumed,plumed2019promoting,invernizzi2020unified,invernizzi2020rethinking}.

WT-MetaD works by adding a bias formed from a history dependent sum of progressively shrinking Gaussian hills \cite{barducci2008well,bussi2020using}. 
The bias at time $t$ for CV value $Q_i$ is given by the expression
\begin{equation}
    V(Q_i,t) = \sum_{\tau<t} h e^{-V(Q_i,\tau)/\Delta T} e^{-\frac{Q(\bm{x}(\tau))-Q_i)^2}{2\sigma^2}},
\end{equation}
where $h$ is the initial hill height, $\sigma$ sets the width of the Gaussians, and $\Delta T$ is an effective sampling temperature for the CVs. Rather than setting $\Delta T$, one typically chooses the bias factor $\gamma=(T+\Delta T)/T$, which sets the smoothness of the sampled distribution \cite{barducci2008well,bussi2020using}.
Asymptotically, a free energy surface (FES) can be estimated from the applied bias by $F(Q)=-\frac{\gamma}{\gamma-1} V(Q,t\rightarrow\infty)$ \cite{dama2014well,bussi2020using} or using a reweighting scheme \cite{tiwary2015time,bussi2020using}.

In contrast to the use of sum of Gaussians in traditional MetaD, OPES-MetaD applies a bias that is based on a kernel density estimate of the probability distribution over the whole space, which is iteratively updated \cite{invernizzi2020unified,invernizzi2020rethinking}. 
The bias at time $t$ for CV value $Q_i$ is given by the expression
\begin{equation}
    V(Q_i) = k_B T \left(\frac{\gamma -1 }{\gamma}\right)  \log\left(\frac{P_t(Q_i)}{Z_t} + \epsilon \right). 
\end{equation}
Here in the prefactor, $T$ is the temperature, $k_B$ is Boltzmann's constant, and $\gamma$ is the bias factor. $P_t(Q)$ is the current estimate of the probability distribution, $Z_t$ is a normalization factor that comes from integrating over sampled $Q$ space. Finally, $\epsilon=\exp(\frac{\Delta E}{k_B T}\frac{\gamma}{\gamma-1})$ is a regularization constant that insures the maximum bias  that can be applied is $\Delta E$.
For one of our systems, we found that limiting the maximum bias using OPES-MetaD helped prevent unphysical exploration along our LDA coordinate (this is also possible using other approaches such as Metabasin Metadynamics \cite{dama2015exploring}). 
Even with this limitation, we apply additional wall potentials to prevent exploration well beyond the LDA values for each of our two states.
As in WT-MetaD, $F(Q)$ can be directly estimated from $V(Q)$ by $F(Q)\approx -\frac{\gamma}{\gamma-1}V(Q)$ or through a reweighting scheme \cite{invernizzi2020rethinking}.
Details of the sampling parameters used for each system are given in Sec.~\ref{sec:sim_details}.

\subsection{Implementation}
Clustering and iterative alignment of trajectory frames prior to learning LDA vectors is performed using our \texttt{shapeGMMTorch} package, which is a high performance re-implementation of the methods from Ref.~\citenum{shapeGMM}, implemented with \texttt{pyTorch} \cite{paszke2019pytorch} for accelerated computation on GPUs. \texttt{shapeGMMTorch} is available from \url{https://github.com/mccullaghlab/shapeGMMTorch} and can easily be installed in python using the command \texttt{pip install shapeGMMTorch}.
We have also created a wrapper library for the training of LDA vectors directly from positional data, which is available from \url{https://github.com/mccullaghlab/pLDA} and which can be easily installed with \texttt{pip install posLDA} (although this wrapper was not used in the analysis performed in this paper as it was not yet available).
Within posLDA, vectors are learned using the SVD implementation of the \texttt{scikit-learn} LinearDiscriminantAnalysis package \cite{scikit-learn}. 

In order to compute and bias these vectors on the fly within MD simulations, the optimal alignment and linear combination procedure has been implemented in the \texttt{PLUMED} open source library \cite{tribello2014plumed,plumed2019promoting}.  
All procedures, analysis for every case studied in this work, and \texttt{PLUMED} code are made available at \url{https://github.com/hocky-research-group/posLDA_paper_2023}, and the code for computing LDA coordinates and Mahalanobis distances on positions will be contributed as a module to \texttt{PLUMED} shortly.

\section{Results and Discussion}
\subsection{LDA is a Good Reaction Coordinate for HP35 Folding}
\begin{figure}[ht!]
    \centering
    \includegraphics{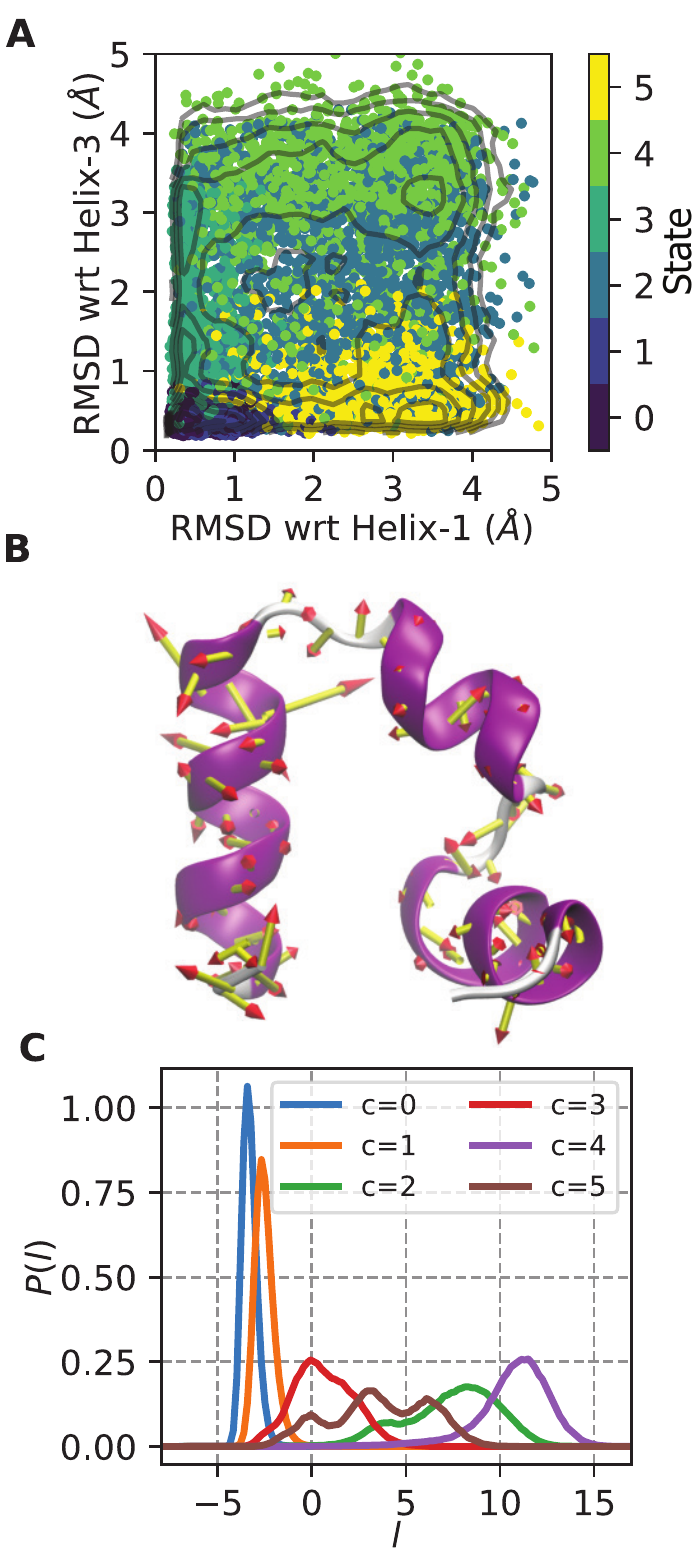}
    \caption{\textbf{Folding/unfolding coordinate for HP35.}  (A) Points from HP35 trajectory are colored by state assignment and mapped into natural folding coordinates of the RMSD of residues in helix 1 or helix 3 to that in the folded state (which is a 3 helix bundle). State 0 is the most folded and state 4 is the most unfolded. Contours shown are every 0.5 kcal/mol in the range (0,6).
    (B) Porcupine plot showing the magnitude of the LDA coefficients trained only on states 0 and 4 from A, overlaid on the starting HP35 structure.
    (C) Histogram of LDA coordinate $l$ for each separate state. $l$ evenly separates all states, with state 0 and 4 at maximum separation.}
    \label{fig:hp35_lda}
\end{figure}
In previous work, we applied our shapeGMM clustering approach to a 305 $\mu$s trajectory of a 35-amino acid fast-folding folding mutant Villin headpiece domain (HP35), obtained from the D.E. Shaw Research Group \cite{Piana2012}.
From our data, we choose to study a six state representation of the data, whose states produce an interpretable representation of folding and unfolding, and which is found not to be overfit by a cross-validation approach. 
The definition of this six state model, $\{\bm{\mu_i},\bm{\Sigma_i}\}_{K=6}$ is trained from 25,000 frames out of $\sim 1.5$ million, and then all frames are assigned to clusters based on which cluster center it is closest to in terms of Mahalanobis distance on positions.

A single folding/unfolding coordinate is constructed by performing LDA on frames assigned to the folded and unfolded states.  The folded and unfolded states were assigned based on the RMSD to folded helix 1 and RMSD to folded helix 2 2D map shown in Fig.\ref{fig:hp35_lda}A for this long trajectory with points colored by the assigned states. 
From this figure, we can assign state 0 as the folded state because it is the state with lowest RMSDs (it also has the largest population) and state 4 as the most unfolded state because it is the state with the largest RMSDs. 
LDA is performed on these two states to produce a single LD vector, denoted $l$, after an iterative alignment of the amalgamated two-state trajectory, as described above. 
The magnitude of the coefficients in this vector are illustrated as particle displacement vectors in the porcupine plot in Fig.~\ref{fig:hp35_lda}B.
Fig.\ref{fig:hp35_lda}C shows a histogram of $l$ values for each state.
We see from this data that this coordinate separates state 0 ($l\approx-3$) and state 4 ($l\approx 12$) . 
To our surprise, this single coordinate, which was trained only on data from state 0 and state 4,  separates the other four states as well, which suggests that it might be sufficient to produce transitions between folded and unfolded through physically meaningful configurations. 

\begin{figure}[ht]
    \centering
    \includegraphics[width=\columnwidth]{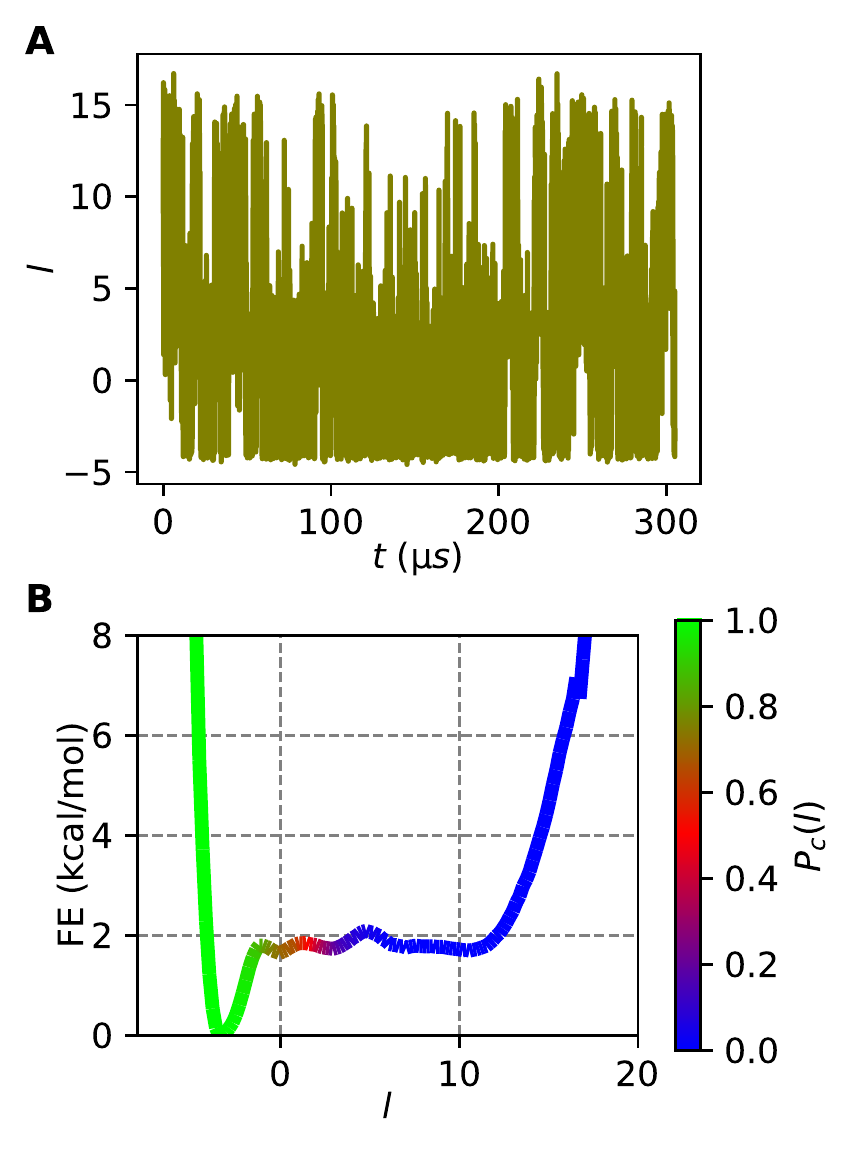}
    \caption{\textbf{LDA results for the folding/unfolding of HP35 from unbiased MD}.  (A) LDA coordinate trained on states 0 and 4 vs. time for the full 305 $\mathrm{\mu}s$ HP35 trajectory shows many transitions between folded ($\sim-3$) and unfolded ($\sim~12)$ states. (B) Free energy vs $l$ for this data, colored by the committor probability in each bin, using 150 bins for the range -8 to 20.}
    \label{fig:hp35_comm}
\end{figure}

Fig.~\ref{fig:hp35_comm}A shows the variation of $l$ versus time for this long trajectory, and exhibits many transitions between the folded ($l\approx -3$) and unfolded ($l\approx 12$) states (for comparison, Ref.~\citenum{Piana2011} found that this long trajectory contains  61 folding transitions with their definition of folding). 
In order to assess the quality of this CV, we compute the committor of each frame in the trajectory $c(\bm{x_t})$ \cite{du1998transition,bolhuis2002transition,ma2005automatic}, which for time $t$ is 1 if the system reaches a folded state before reaching an unfolded state in the times following $t$.

To assess the quality of a reaction coordinate, we can compute the committor probability for each value of $l$ on a grid of size $\delta l$.
\begin{eqnarray}
    P_c(l_i) = \frac{1}{M_i} \sum_{t=1}^{N_\mathrm{frames}} c(\bm{x_t})[l(\bm{x_t})\in (l_i-\delta l,l_i+\delta l)] \\
    M_i = \sum_{t=1}^{N_\mathrm{frames}}[l(\bm{x_t})\in (l_i-\delta l,l_i+\delta l)].
\end{eqnarray}

In Fig.~\ref{fig:hp35_comm}B, we show the approximate FES along $l$ computed as $F(l)=-k_B T \ln P(l)$ for the long unbiased trajectory, colored by the value of $P_c(l)$. 
The FES shows a stable well at a value of $l=-3$ corresponding to the highest population state, the folded one, and very shallow minima for each of the other states. 
The value of $P_c$ varies continuously from 1 to 0 along this coordinate, reaching a value of 0.5 at $l=1$, just outside the folded basin. 
By this metric, our very simple CV is a good CV for characterizing the transition between folded and unfolded states, although the lack of a high barrier separating the two states (due to the system being near its melting temperature) makes it more ambiguous how close the point of $P_c=0.5$ is to a classic transition state.
The coincidence of $P_c=0.5$ with a clear barrier is observed in Fig.~\ref{fig:compare_hp35_2state_6state} where we train using all 6 states, but for this paper we chose to focus only on one dimensional LDA spaces.

\subsection{LDA is a Reasonable Sampling Coordinate for HP35 Folding}
\begin{figure}[ht]
    \centering
    \includegraphics[width=\columnwidth]{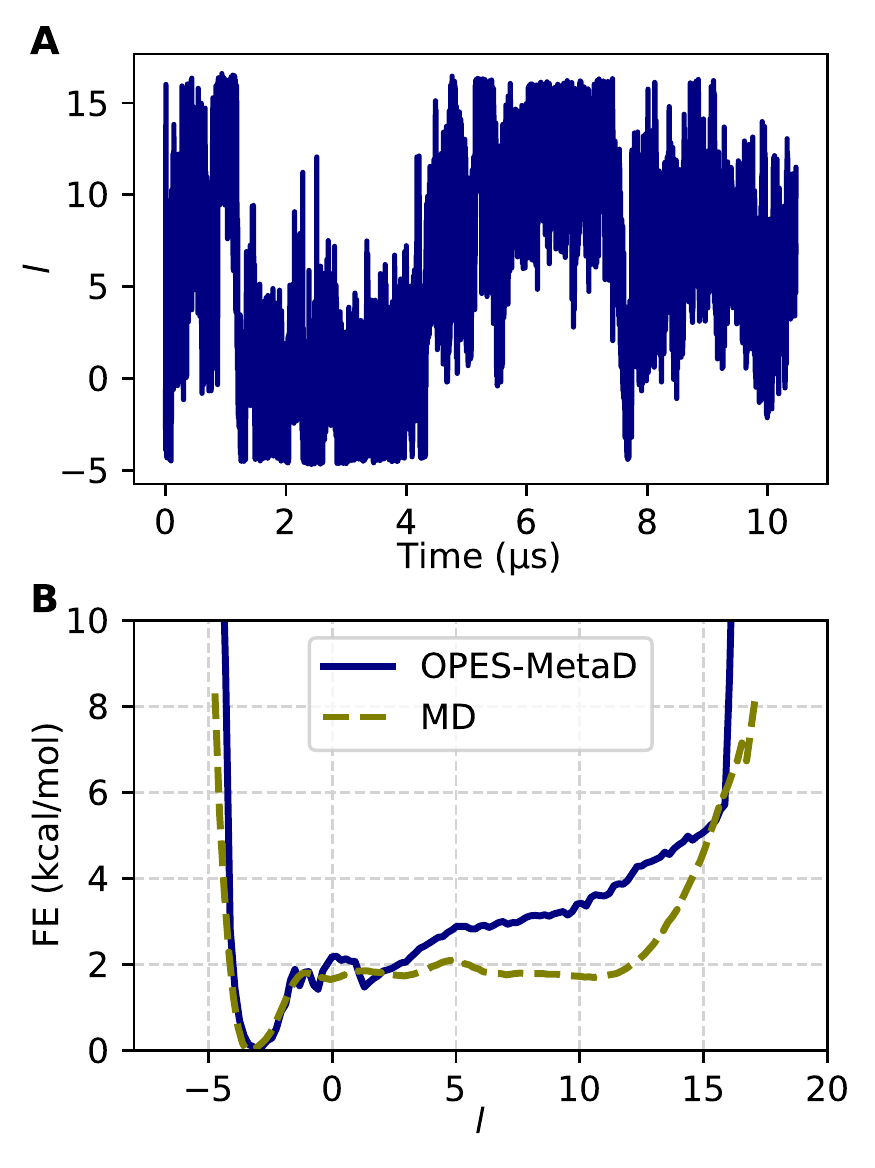}
    \caption{\textbf{OPES-MetaD sampling on HP35 using the folding/unfolding LDA coordinate. } (A) LDA coordinate vs. times for OPES-MetaD simulation. (B) Comparison of free energy estimated from unbiased MD and OPES-MetaD. }
    \label{fig:hp35_opes_fes}
\end{figure}

To assess the ability to sample along an LDA coordinate we perform OPES-MetaD to bias the system to explore $l$. 
For the MetaD parameters listed in Sec.~\ref{sec:sim_details}, we find that transitions between the folded and unfolded state are accelerated. 
For these parameters, we are able to obtain several transitions in 10 $\mathrm{\mu} s$, resulting in a estimated FES that is in fair agreement with that obtained from the long unbiased trajectory consider it is obtained in only 3\% of the MD time. 
In Fig.~\ref{fig:hp35_fe_rmsd} we show the FES projected on natural folding coordinates, and see that our sampling does a good job capturing the main features of the long unbiased trajectory, including the presence of intermediates along the x- and y- axes, and the high energy unfolded state located in the upper right. 
However, the most unfolded regions are unexplored and the statistical weight of the central intermediate basin is incorrect. 
Shorter replicates of simulations starting from different initial structures (Fig.~\ref{fig:hp35_compare_four_runs}) show the variance in FES estimates that could arise if one is not careful to converge sampling.
On the whole, our results show that our simple LDA coordinate is a promising first step for sampling between two states of a complex biomolecule.

\subsection{Accurate Sampling Using LDA for a Bistable Helix}
The LDA procedure can be applied to determine a reaction coordinate separating two states even without sampling the actual transition (analogous to Ref.~\citenum{mendels2018collective}).
To assess this behavior we investigate the right to left-handed helix transition of (Aib)$_9$, a nine residue peptide formed from the achiral $\alpha$-aminoisobutyryl amino acid \cite{karle1990structural}. 
The helical states of achiral molecules must by symmetry have equal free energy, and we previously took advantage of this property in benchmarking sampling and clustering methods \cite{Hartmann2020,shapeGMM}. 
The properties of (Aib)$_9$ have been characterized in simulation including recently as a tool to benchmark advanced methods for RC optimization  \cite{buchenberg2015hierarchical,biswas2018metadynamics,mehdi2022accelerating}.

\begin{figure}[ht]
    \centering
    \includegraphics[width=\columnwidth]{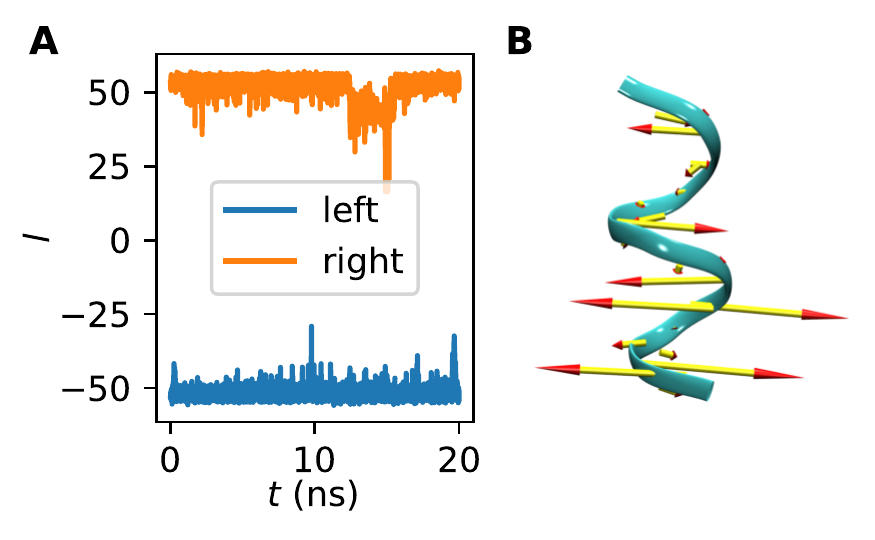}
    \caption{\textbf{LDA coordinate for helical inversion of (Aib)$_9$}.  (A) LDA coordinate $l$ vs. time for training data starting from left and right handed helix. (B) Porcupine plot showing the magnitude of the LDA coefficients on the left handed helical structure.}
    \label{fig:aib9_training}
\end{figure}

We performed 20 ns simulations starting from the left and right handed states of (Aib)$_9$ using inputs from Ref.~\citenum{mehdi2022accelerating} (see Sec.~\ref{sec:sim_details} for details).
We then performed a iterative alignment of these data to compute a global $(\bm{\mu},\bm{\Sigma})$, and then computed a single LDA vector between those frames coming from the left- and right- states, respectively. 
Fig.~\ref{fig:aib9_training} shows that this coordinate separates the training data with $l\sim50$ indicating a right-handed helix and $l\sim-50$ indicating a left-handed helix.
The left-handed helix is the starting point for further runs.

\begin{figure}[ht]
    \centering
    \includegraphics[width=\columnwidth]{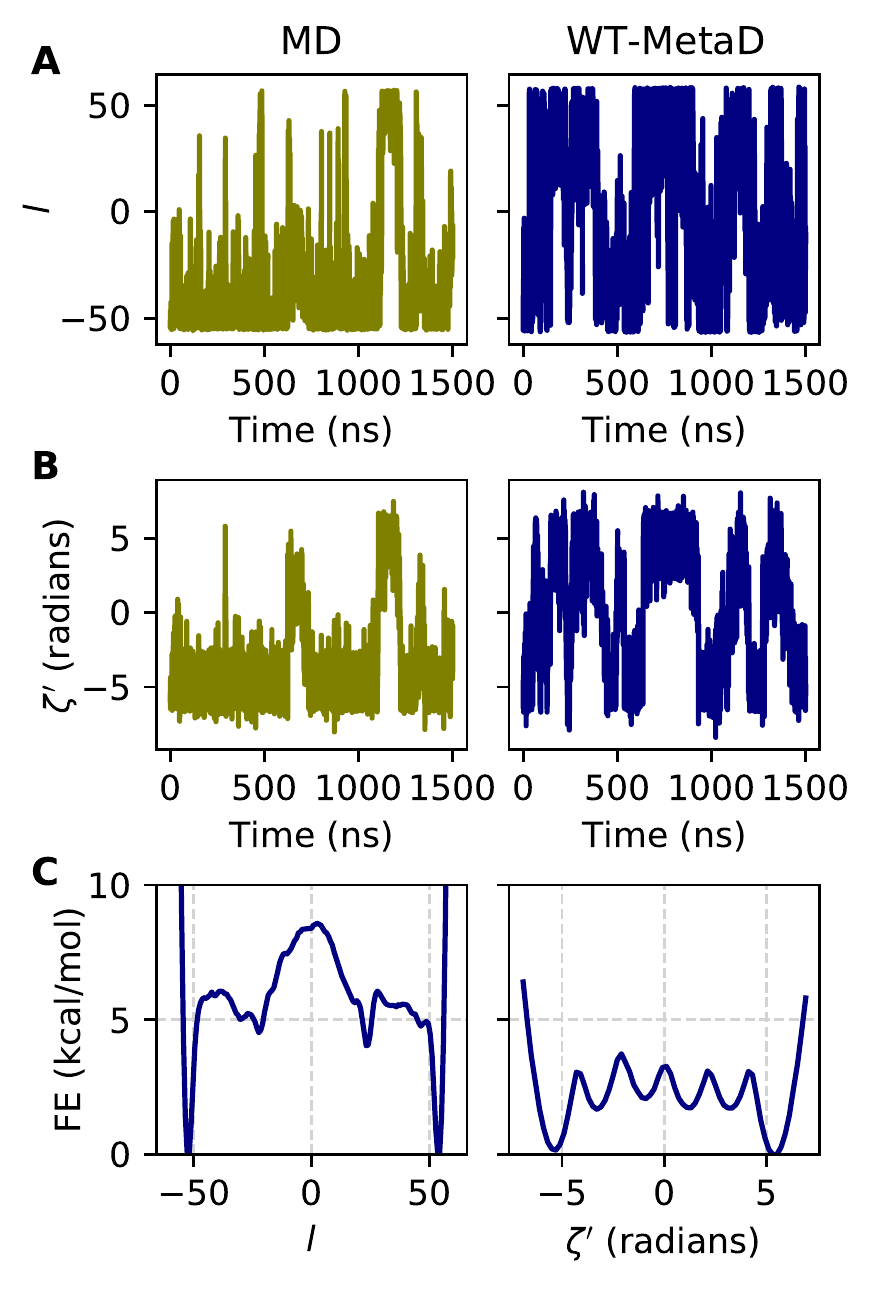}
    \caption{\textbf{Metadynamics sampling results along the LDA coordinate for (Aib)$_9$}. (A) LDA coordinate $l$ vs. time for 1.5 $\mathrm{\mu}$s of conventional MD and WT-MetaD. (B) Helical parameter $\zeta$ vs. time for the trajectories in A. (C) FES along $l$ and $\zeta$ from WT-MetaD simulations. }
    \label{fig:aib9_summary}
\end{figure}
Having trained $l$, we next performed conventional and WT-MetaD simulations starting from the structure in Fig.~\ref{fig:aib9_training}A. 
Fig.~\ref{fig:aib9_summary}A shows that MetaD (right) substantially increases the rate of transition between the left and right handed states as compared to conventional MD (left).

A more chemically motivated way of computing the helicity of (Aib)$_9$ is the parameter $\zeta'=-\sum_{n=3}^7 $$\phi_n$, the negative sum over the central backbone $\phi$ dihedral angles \cite{mehdi2022accelerating}. This quantity takes on the value approximately 5 for right handed and -5 for left handed helices \cite{mehdi2022accelerating}. Fig.~\ref{fig:aib9_summary}B shows qualitatively similar behavior for $\zeta'$ as $l$.

Fig.~\ref{fig:aib9_summary}C shows the FES computed for these two quantities. 
The sampled $l$ has a nearly perfectly symmetrical FES, and in particular the free energy difference between the left and right handed states is negligible.
For the FES of the non-biased $\zeta'$ computed by reweighting, the result is nearly as symmetrical, and the offset in free energy between the left and right handed size is visible but minuscule.
This result appears to be as good as that obtained in Ref.~\citenum{mehdi2022accelerating}, which uses a very sophisticated iterative process and 900 ns of unbiased and biased simulation data to obtain an optimized sampling coordinate as compared to our 40 ns of input data; however, their optimized coordinate appears to perform better in terms of transitions per unit time generated with their choice of MetaD parameters.

\section{Conclusions and outlook}
In this work, we demonstrated that LDA on positions computed from two states of a system produces a good reaction coordinate, both in terms of state transition kinetics and our ability to bias that coordinate to assess the FES along that coordinate.
This was true for (Aib)$_9$ even though the RC was trained only using short simulations starting in each state, making this a promising approach even when only structures of  endpoints of a process are available.
In contrast to Ref.~\citenum{mendels2018collective} where input features were internal coordinates, we were able to use standard LDA rather than HLDA in this case and achieve good performance.
We note that LDA on positions would not apply directly to the case of molecular dissociation since the dissociated states cannot be aligned to a single average structure; however, we do think this coordinate would work well for apo-holo transitions of a biomolecule and could easily be combined with a ligand-distance coordinate for that application.

For HP35, multidimensional LDA by construction better separates all of the states of the molecule and may also provide an even better reaction coordinate for kinetics (Fig.~\ref{fig:compare_hp35_2state_6state}), and we will explore whether this can also improve our sampling results going forward.
However, this is not an option when information about multiple states is unavailable a priori (such as in the case of (Aib)$_9$) which is why we did not include it here.
For cases like that, it would be intriguing to first sample along the 1-dimensional reaction coordinate, then train a GMM with a higher number of states, and continue iterating this approach.

The use of states defined from our GMM clustering approach presents both an advantage and disadvantage as illustrated in the case of HP35. 
Our approach allowed us to explore the folding/unfolding process and most of the conformational landscape (Fig.~\ref{fig:hp35_fe_rmsd}), but we were not able to accurately sample the FES around the unfolded state. 
For fully sampling a broad and entropy dominated state, use of temperature accelerated methods or combining our approach with tempering may provide more accurate information in this region.

In both the case of HP35 and (Aib)$_9$, we were able to accelerate transitions between two states using MetaD or OPES-MetaD. 
In our hands, the biased simulations were sensitive to sampling protocol in terms of being able to run microseconds or longer without ``crashing''. 
HP35 was less sensitive to this issue using OPES-MetaD, while (Aib)$_9$ performed better with standard WT-MetaD. 
For this reason, we used small bias factors and hill heights/barrier heights, which resulted in fewer transitions and presumably worse convergence in fixed simulation time.
We speculate that some of this sensitivity may come from our choice of the global trajectory mean and covariance as the reference state when computing our LDA vectors.
If the reference state is hard to align to, then the value of $l$ will be very sensitive changes in molecular configuration, and that could lead to instability. 
An alternative approach would be to take one of the two states as the reference, which we will explore further in the future.
However, this has the similar disadvantage that structures in the other state may be very difficult to align to this one if they are quite different.
A compelling option is presented in the ATLAS method of Ref.~\citenum{giberti2021global}, where bias is computed along vectors to multiple reference states, weighted by distance from that reference state, and we plan to assess that approach going forward.

\section{Simulation Details}
\label{sec:sim_details}
All simulations were performed using GROMACS 2019.6 \cite{abraham2015gromacs}  with PLUMED 2.9.0-dev \cite{tribello2014plumed,plumed2019promoting}.
GROMACS `mdp' parameter files and PLUMED input files are available in our paper's github repository for complete details.

\subsection{HP35 Simulations}
A 305 $\mu$s all-atom simulation of Nle/Nle HP35 at $T=360K$ from Piana et al.\cite{Piana2012} was analyzed.
The simulation was performed using the Amber ff99SB*-ILDN force field and TIP3P water model.
In that simulation, protein configurations were saved every 200 ps, for a total of $\sim$1.5M frames.
For our simulations, we solvate and equilibrate a fresh system using the same forcefield at 40mM NaCl. 
Minimization and equilibration are performed using a standard protocol\footnote{\url{http://www.mdtutorials.com/gmx/lysozyme/index.html}}, at which point NPT simulations are initiated at $T=360K$. mdp files for all steps of this procedure and the topology files are all available in the paper's github.

OPES-MetaD simulations are performed with $\gamma=8$, $\Delta E=10$ kcal/mol, pace of 500 steps, and a multiple time step \cite{ferrarotti2015accurate} stride of 2.
Quadratic walls are applied at $l=5$ and $l=-15$ with a bias coefficient of 125 kcal/mol/\AA$^2$.

\subsection{(Aib)9 Simulations}
Equilibrated inputs for (Aib)$_9$ were provided by the authors of Ref.~\citenum{mehdi2022accelerating}. 
In brief, simulations  using the CHARMM36m forcefield and TIP3P water \cite{huang2017charmm36m}.
MD simulations are performed in NPT with a 2 fs timestep at $T=400K$. 

WT-MetaD simulations are performed with $h=0.005$ kcal/mol, $\sigma=0.43$, $\gamma=2$ and a multiple time step \cite{ferrarotti2015accurate} stride of 2.
Quadratic walls are applied at $l=70$ and $l=-60$ with a bias coefficient of 125 kcal/mol/\AA$^2$.
$\sigma$ was chosen as the $\sigma_l/3$ where $\sigma_l$ was the standard deviation in $l$ over the 20 ns simulation starting from the left helical state used in the training of the CV. 

\begin{acknowledgement}
We thank the D.E. Shaw Research for providing simulation data on the HP35 protein and we thank the Tiwary lab for providing their input files for (Aib)$_9$.
SS and GMH were supported by the National Institutes of Health through the award R35GM138312. SS was also partially supported by a graduate fellowship from the Simons Center for Computational Physical Chemistry (SCCPC) at NYU (SF Grant No. 839534).
MM would like to acknowledge funding from National Institute of Allergy and Infectious Diseases of the National Institutes of Health under award number R01AI166050.
This work was supported in part through the NYU IT High Performance Computing resources, services, and staff expertise, and simulations were partially executed on resources supported by the SCCPC at NYU.
\end{acknowledgement}

\bibliography{refs}

\providecommand{\latin}[1]{#1}
\makeatletter
\providecommand{\doi}
  {\begingroup\let\do\@makeother\dospecials
  \catcode`\{=1 \catcode`\}=2 \doi@aux}
\providecommand{\doi@aux}[1]{\endgroup\texttt{#1}}
\makeatother
\providecommand*\mcitethebibliography{\thebibliography}
\csname @ifundefined\endcsname{endmcitethebibliography}
  {\let\endmcitethebibliography\endthebibliography}{}
\begin{mcitethebibliography}{54}
\providecommand*\natexlab[1]{#1}
\providecommand*\mciteSetBstSublistMode[1]{}
\providecommand*\mciteSetBstMaxWidthForm[2]{}
\providecommand*\mciteBstWouldAddEndPuncttrue
  {\def\EndOfBibitem{\unskip.}}
\providecommand*\mciteBstWouldAddEndPunctfalse
  {\let\EndOfBibitem\relax}
\providecommand*\mciteSetBstMidEndSepPunct[3]{}
\providecommand*\mciteSetBstSublistLabelBeginEnd[3]{}
\providecommand*\EndOfBibitem{}
\mciteSetBstSublistMode{f}
\mciteSetBstMaxWidthForm{subitem}{(\alph{mcitesubitemcount})}
\mciteSetBstSublistLabelBeginEnd
  {\mcitemaxwidthsubitemform\space}
  {\relax}
  {\relax}

\bibitem[H{\'e}nin \latin{et~al.}(2022)H{\'e}nin, Leli{\`e}vre, Shirts,
  Valsson, and Delemotte]{henin2022enhanced}
H{\'e}nin,~J.; Leli{\`e}vre,~T.; Shirts,~M.; Valsson,~O.; Delemotte,~L.
  Enhanced Sampling Methods for Molecular Dynamics Simulations [Article v1. 0].
  \emph{LiveCoMS} \textbf{2022}, \emph{4}, 1583--1583\relax
\mciteBstWouldAddEndPuncttrue
\mciteSetBstMidEndSepPunct{\mcitedefaultmidpunct}
{\mcitedefaultendpunct}{\mcitedefaultseppunct}\relax
\EndOfBibitem
\bibitem[Ma and Dinner(2005)Ma, and Dinner]{ma2005automatic}
Ma,~A.; Dinner,~A.~R. Automatic method for identifying reaction coordinates in
  complex systems. \emph{J. Phys. Chem. B} \textbf{2005}, \emph{109},
  6769--6779\relax
\mciteBstWouldAddEndPuncttrue
\mciteSetBstMidEndSepPunct{\mcitedefaultmidpunct}
{\mcitedefaultendpunct}{\mcitedefaultseppunct}\relax
\EndOfBibitem
\bibitem[Hashemian \latin{et~al.}(2013)Hashemian, Mill{\'a}n, and
  Arroyo]{hashemian2013modeling}
Hashemian,~B.; Mill{\'a}n,~D.; Arroyo,~M. Modeling and enhanced sampling of
  molecular systems with smooth and nonlinear data-driven collective variables.
  \emph{J. Chem. Phys.} \textbf{2013}, \emph{139}, 214101\relax
\mciteBstWouldAddEndPuncttrue
\mciteSetBstMidEndSepPunct{\mcitedefaultmidpunct}
{\mcitedefaultendpunct}{\mcitedefaultseppunct}\relax
\EndOfBibitem
\bibitem[Tiwary and Berne(2016)Tiwary, and Berne]{tiwary2016spectral}
Tiwary,~P.; Berne,~B. Spectral gap optimization of order parameters for
  sampling complex molecular systems. \emph{Proc. Natl. Acad. Sci.}
  \textbf{2016}, \emph{113}, 2839--2844\relax
\mciteBstWouldAddEndPuncttrue
\mciteSetBstMidEndSepPunct{\mcitedefaultmidpunct}
{\mcitedefaultendpunct}{\mcitedefaultseppunct}\relax
\EndOfBibitem
\bibitem[Chen and Ferguson(2018)Chen, and Ferguson]{chen2018molecular}
Chen,~W.; Ferguson,~A.~L. Molecular enhanced sampling with autoencoders:
  On-the-fly collective variable discovery and accelerated free energy
  landscape exploration. \emph{J. Comp. Chem.} \textbf{2018}, \emph{39},
  2079--2102\relax
\mciteBstWouldAddEndPuncttrue
\mciteSetBstMidEndSepPunct{\mcitedefaultmidpunct}
{\mcitedefaultendpunct}{\mcitedefaultseppunct}\relax
\EndOfBibitem
\bibitem[Mendels \latin{et~al.}(2018)Mendels, Piccini, and
  Parrinello]{mendels2018collective}
Mendels,~D.; Piccini,~G.; Parrinello,~M. Collective variables from local
  fluctuations. \emph{J. Phys. Chem. Lett.} \textbf{2018}, \emph{9},
  2776--2781\relax
\mciteBstWouldAddEndPuncttrue
\mciteSetBstMidEndSepPunct{\mcitedefaultmidpunct}
{\mcitedefaultendpunct}{\mcitedefaultseppunct}\relax
\EndOfBibitem
\bibitem[Mendels \latin{et~al.}(2018)Mendels, Piccini, Brotzakis, Yang, and
  Parrinello]{mendels2018folding}
Mendels,~D.; Piccini,~G.; Brotzakis,~Z.~F.; Yang,~Y.~I.; Parrinello,~M. Folding
  a small protein using harmonic linear discriminant analysis. \emph{J. Chem.
  Phys.} \textbf{2018}, \emph{149}, 194113\relax
\mciteBstWouldAddEndPuncttrue
\mciteSetBstMidEndSepPunct{\mcitedefaultmidpunct}
{\mcitedefaultendpunct}{\mcitedefaultseppunct}\relax
\EndOfBibitem
\bibitem[Wehmeyer and No{\'e}(2018)Wehmeyer, and No{\'e}]{wehmeyer2018time}
Wehmeyer,~C.; No{\'e},~F. Time-lagged autoencoders: Deep learning of slow
  collective variables for molecular kinetics. \emph{J. Chem. Phys.}
  \textbf{2018}, \emph{148}, 241703\relax
\mciteBstWouldAddEndPuncttrue
\mciteSetBstMidEndSepPunct{\mcitedefaultmidpunct}
{\mcitedefaultendpunct}{\mcitedefaultseppunct}\relax
\EndOfBibitem
\bibitem[Piccini \latin{et~al.}(2018)Piccini, Mendels, and
  Parrinello]{piccini2018metadynamics}
Piccini,~G.; Mendels,~D.; Parrinello,~M. Metadynamics with discriminants: A
  tool for understanding chemistry. \emph{J. Chem. Theory Comput.}
  \textbf{2018}, \emph{14}, 5040--5044\relax
\mciteBstWouldAddEndPuncttrue
\mciteSetBstMidEndSepPunct{\mcitedefaultmidpunct}
{\mcitedefaultendpunct}{\mcitedefaultseppunct}\relax
\EndOfBibitem
\bibitem[Sultan and Pande(2018)Sultan, and Pande]{sultan2018automated}
Sultan,~M.~M.; Pande,~V.~S. Automated design of collective variables using
  supervised machine learning. \emph{J. Chem. Phys.} \textbf{2018}, \emph{149},
  094106\relax
\mciteBstWouldAddEndPuncttrue
\mciteSetBstMidEndSepPunct{\mcitedefaultmidpunct}
{\mcitedefaultendpunct}{\mcitedefaultseppunct}\relax
\EndOfBibitem
\bibitem[Ribeiro \latin{et~al.}(2018)Ribeiro, Bravo, Wang, and
  Tiwary]{ribeiro2018reweighted}
Ribeiro,~J. M.~L.; Bravo,~P.; Wang,~Y.; Tiwary,~P. Reweighted autoencoded
  variational Bayes for enhanced sampling (RAVE). \emph{J. Chem. Phys.}
  \textbf{2018}, \emph{149}, 072301\relax
\mciteBstWouldAddEndPuncttrue
\mciteSetBstMidEndSepPunct{\mcitedefaultmidpunct}
{\mcitedefaultendpunct}{\mcitedefaultseppunct}\relax
\EndOfBibitem
\bibitem[Zhang \latin{et~al.}(2019)Zhang, Niu, Piccini, Mendels, and
  Parrinello]{zhang2019improving}
Zhang,~Y.-Y.; Niu,~H.; Piccini,~G.; Mendels,~D.; Parrinello,~M. Improving
  collective variables: The case of crystallization. \emph{J. Chem. Phys.}
  \textbf{2019}, \emph{150}, 094509\relax
\mciteBstWouldAddEndPuncttrue
\mciteSetBstMidEndSepPunct{\mcitedefaultmidpunct}
{\mcitedefaultendpunct}{\mcitedefaultseppunct}\relax
\EndOfBibitem
\bibitem[Wang \latin{et~al.}(2020)Wang, Ribeiro, and Tiwary]{wang2020machine}
Wang,~Y.; Ribeiro,~J. M.~L.; Tiwary,~P. Machine learning approaches for
  analyzing and enhancing molecular dynamics simulations. \emph{Curr. Opin.
  Struct. Biol.} \textbf{2020}, \emph{61}, 139--145\relax
\mciteBstWouldAddEndPuncttrue
\mciteSetBstMidEndSepPunct{\mcitedefaultmidpunct}
{\mcitedefaultendpunct}{\mcitedefaultseppunct}\relax
\EndOfBibitem
\bibitem[No{\'e} \latin{et~al.}(2020)No{\'e}, Tkatchenko, M{\"u}ller, and
  Clementi]{noe2020machine}
No{\'e},~F.; Tkatchenko,~A.; M{\"u}ller,~K.-R.; Clementi,~C. Machine learning
  for molecular simulation. \emph{Annu. Rev. Phys. Chem.} \textbf{2020},
  \emph{71}, 361--390\relax
\mciteBstWouldAddEndPuncttrue
\mciteSetBstMidEndSepPunct{\mcitedefaultmidpunct}
{\mcitedefaultendpunct}{\mcitedefaultseppunct}\relax
\EndOfBibitem
\bibitem[Sidky \latin{et~al.}(2020)Sidky, Chen, and Ferguson]{sidky2020machine}
Sidky,~H.; Chen,~W.; Ferguson,~A.~L. Machine learning for collective variable
  discovery and enhanced sampling in biomolecular simulation. \emph{Mol. Phys.}
  \textbf{2020}, \emph{118}, e1737742\relax
\mciteBstWouldAddEndPuncttrue
\mciteSetBstMidEndSepPunct{\mcitedefaultmidpunct}
{\mcitedefaultendpunct}{\mcitedefaultseppunct}\relax
\EndOfBibitem
\bibitem[Bonati \latin{et~al.}(2020)Bonati, Rizzi, and
  Parrinello]{bonati2020data}
Bonati,~L.; Rizzi,~V.; Parrinello,~M. Data-driven collective variables for
  enhanced sampling. \emph{J. Chem. Phys. Lett.} \textbf{2020}, \emph{11},
  2998--3004\relax
\mciteBstWouldAddEndPuncttrue
\mciteSetBstMidEndSepPunct{\mcitedefaultmidpunct}
{\mcitedefaultendpunct}{\mcitedefaultseppunct}\relax
\EndOfBibitem
\bibitem[Karmakar \latin{et~al.}(2021)Karmakar, Invernizzi, Rizzi, and
  Parrinello]{karmakar2021collective}
Karmakar,~T.; Invernizzi,~M.; Rizzi,~V.; Parrinello,~M. Collective variables
  for the study of crystallisation. \emph{Mol. Phys.} \textbf{2021},
  \emph{119}, e1893848\relax
\mciteBstWouldAddEndPuncttrue
\mciteSetBstMidEndSepPunct{\mcitedefaultmidpunct}
{\mcitedefaultendpunct}{\mcitedefaultseppunct}\relax
\EndOfBibitem
\bibitem[Tsai \latin{et~al.}(2021)Tsai, Smith, and Tiwary]{tsai2021sgoop}
Tsai,~S.-T.; Smith,~Z.; Tiwary,~P. Sgoop-d: Estimating kinetic distances and
  reaction coordinate dimensionality for rare event systems from
  biased/unbiased simulations. \emph{J. Chem. Theory Comput.} \textbf{2021},
  \emph{17}, 6757--6765\relax
\mciteBstWouldAddEndPuncttrue
\mciteSetBstMidEndSepPunct{\mcitedefaultmidpunct}
{\mcitedefaultendpunct}{\mcitedefaultseppunct}\relax
\EndOfBibitem
\bibitem[Hooft \latin{et~al.}(2021)Hooft, Perez~de Alba~Ortiz, and
  Ensing]{hooft2021discovering}
Hooft,~F.; Perez~de Alba~Ortiz,~A.; Ensing,~B. Discovering collective variables
  of molecular transitions via genetic algorithms and neural networks. \emph{J.
  Chem. Theory Comput.} \textbf{2021}, \emph{17}, 2294--2306\relax
\mciteBstWouldAddEndPuncttrue
\mciteSetBstMidEndSepPunct{\mcitedefaultmidpunct}
{\mcitedefaultendpunct}{\mcitedefaultseppunct}\relax
\EndOfBibitem
\bibitem[Sun \latin{et~al.}(2022)Sun, Vandermause, Batzner, Xie, Clark, Chen,
  and Kozinsky]{sun2022multitask}
Sun,~L.; Vandermause,~J.; Batzner,~S.; Xie,~Y.; Clark,~D.; Chen,~W.;
  Kozinsky,~B. Multitask Machine Learning of Collective Variables for Enhanced
  Sampling of Rare Events. \emph{J. Chem. Theory Comput.} \textbf{2022},
  \emph{18}, 2341--2353\relax
\mciteBstWouldAddEndPuncttrue
\mciteSetBstMidEndSepPunct{\mcitedefaultmidpunct}
{\mcitedefaultendpunct}{\mcitedefaultseppunct}\relax
\EndOfBibitem
\bibitem[Rydzewski \latin{et~al.}(2022)Rydzewski, Chen, Ghosh, and
  Valsson]{rydzewski2022reweighted}
Rydzewski,~J.; Chen,~M.; Ghosh,~T.~K.; Valsson,~O. Reweighted Manifold Learning
  of Collective Variables from Enhanced Sampling Simulations. \emph{J. Chem.
  Theory Comput.} \textbf{2022}, \emph{18}, 7179--7192\relax
\mciteBstWouldAddEndPuncttrue
\mciteSetBstMidEndSepPunct{\mcitedefaultmidpunct}
{\mcitedefaultendpunct}{\mcitedefaultseppunct}\relax
\EndOfBibitem
\bibitem[Chen \latin{et~al.}(2019)Chen, Sidky, and Ferguson]{chen2019nonlinear}
Chen,~W.; Sidky,~H.; Ferguson,~A.~L. Nonlinear discovery of slow molecular
  modes using state-free reversible VAMPnets. \emph{J. Chem. Phys.}
  \textbf{2019}, \emph{150}, 214114\relax
\mciteBstWouldAddEndPuncttrue
\mciteSetBstMidEndSepPunct{\mcitedefaultmidpunct}
{\mcitedefaultendpunct}{\mcitedefaultseppunct}\relax
\EndOfBibitem
\bibitem[Bonati \latin{et~al.}(2021)Bonati, Piccini, and
  Parrinello]{bonati2021deep}
Bonati,~L.; Piccini,~G.; Parrinello,~M. Deep learning the slow modes for rare
  events sampling. \emph{Proceedings of the National Academy of Sciences}
  \textbf{2021}, \emph{118}, e2113533118\relax
\mciteBstWouldAddEndPuncttrue
\mciteSetBstMidEndSepPunct{\mcitedefaultmidpunct}
{\mcitedefaultendpunct}{\mcitedefaultseppunct}\relax
\EndOfBibitem
\bibitem[Mehdi \latin{et~al.}(2022)Mehdi, Wang, Pant, and
  Tiwary]{mehdi2022accelerating}
Mehdi,~S.; Wang,~D.; Pant,~S.; Tiwary,~P. Accelerating all-atom simulations and
  gaining mechanistic understanding of biophysical systems through state
  predictive information bottleneck. \emph{J. Chem. Theory Comput.}
  \textbf{2022}, \emph{18}, 3231--3238\relax
\mciteBstWouldAddEndPuncttrue
\mciteSetBstMidEndSepPunct{\mcitedefaultmidpunct}
{\mcitedefaultendpunct}{\mcitedefaultseppunct}\relax
\EndOfBibitem
\bibitem[Klem \latin{et~al.}(2022)Klem, Hocky, and McCullagh]{shapeGMM}
Klem,~H.; Hocky,~G.~M.; McCullagh,~M. Size-and-Shape Space Gaussian Mixture
  Models for Structural Clustering of Molecular Dynamics Trajectories. \emph{J.
  Chem. Theory Comput.} \textbf{2022}, \emph{18}, 3218--3230, PMID:
  35483073\relax
\mciteBstWouldAddEndPuncttrue
\mciteSetBstMidEndSepPunct{\mcitedefaultmidpunct}
{\mcitedefaultendpunct}{\mcitedefaultseppunct}\relax
\EndOfBibitem
\bibitem[Tribello \latin{et~al.}(2010)Tribello, Ceriotti, and
  Parrinello]{tribello2010self}
Tribello,~G.~A.; Ceriotti,~M.; Parrinello,~M. A self-learning algorithm for
  biased molecular dynamics. \emph{Proc. Natl. Acad. Sci.} \textbf{2010},
  \emph{107}, 17509--17514\relax
\mciteBstWouldAddEndPuncttrue
\mciteSetBstMidEndSepPunct{\mcitedefaultmidpunct}
{\mcitedefaultendpunct}{\mcitedefaultseppunct}\relax
\EndOfBibitem
\bibitem[Westerlund and Delemotte(2019)Westerlund, and Delemotte]{InfleCS}
Westerlund,~A.~M.; Delemotte,~L. InfleCS: Clustering Free Energy Landscapes
  with Gaussian Mixtures. \emph{J. Chem. Theory Comput.} \textbf{2019},
  \emph{15}, 6752--6759\relax
\mciteBstWouldAddEndPuncttrue
\mciteSetBstMidEndSepPunct{\mcitedefaultmidpunct}
{\mcitedefaultendpunct}{\mcitedefaultseppunct}\relax
\EndOfBibitem
\bibitem[Giberti \latin{et~al.}(2021)Giberti, Tribello, and
  Ceriotti]{giberti2021global}
Giberti,~F.; Tribello,~G.; Ceriotti,~M. Global free-energy landscapes as a
  smoothly joined collection of local maps. \emph{J. Chem. Theory Comput.}
  \textbf{2021}, \emph{17}, 3292--3308\relax
\mciteBstWouldAddEndPuncttrue
\mciteSetBstMidEndSepPunct{\mcitedefaultmidpunct}
{\mcitedefaultendpunct}{\mcitedefaultseppunct}\relax
\EndOfBibitem
\bibitem[Dryden and Mardia(1998)Dryden, and Mardia]{Dryden1998}
Dryden,~I.~L.; Mardia,~K.~V. \emph{Statistical Shape Analysis}; John Wiley \&
  Sons: Chichester, 1998\relax
\mciteBstWouldAddEndPuncttrue
\mciteSetBstMidEndSepPunct{\mcitedefaultmidpunct}
{\mcitedefaultendpunct}{\mcitedefaultseppunct}\relax
\EndOfBibitem
\bibitem[Bishop and Nasrabadi(2006)Bishop, and Nasrabadi]{bishop2006pattern}
Bishop,~C.~M.; Nasrabadi,~N.~M. \emph{Pattern recognition and machine
  learning}; Springer, 2006; Vol.~4\relax
\mciteBstWouldAddEndPuncttrue
\mciteSetBstMidEndSepPunct{\mcitedefaultmidpunct}
{\mcitedefaultendpunct}{\mcitedefaultseppunct}\relax
\EndOfBibitem
\bibitem[Howland \latin{et~al.}(2003)Howland, Jeon, and Park]{LDA_SVD}
Howland,~P.; Jeon,~M.; Park,~H. Structure Preserving Dimension Reduction for
  Clustered Text Data Based on the Generalized Singular Value Decomposition.
  \emph{SIAM J. Matrix Anal. Appl.} \textbf{2003}, \emph{25}, 165--179\relax
\mciteBstWouldAddEndPuncttrue
\mciteSetBstMidEndSepPunct{\mcitedefaultmidpunct}
{\mcitedefaultendpunct}{\mcitedefaultseppunct}\relax
\EndOfBibitem
\bibitem[Tribello \latin{et~al.}(2014)Tribello, Bonomi, Branduardi, Camilloni,
  and Bussi]{tribello2014plumed}
Tribello,~G.~A.; Bonomi,~M.; Branduardi,~D.; Camilloni,~C.; Bussi,~G. PLUMED 2:
  New feathers for an old bird. \emph{Comp. Phys. Comm.} \textbf{2014},
  \emph{185}, 604--613\relax
\mciteBstWouldAddEndPuncttrue
\mciteSetBstMidEndSepPunct{\mcitedefaultmidpunct}
{\mcitedefaultendpunct}{\mcitedefaultseppunct}\relax
\EndOfBibitem
\bibitem[plu(2019)]{plumed2019promoting}
Promoting transparency and reproducibility in enhanced molecular simulations.
  \emph{Nat. Methods} \textbf{2019}, \emph{16}, 670--673\relax
\mciteBstWouldAddEndPuncttrue
\mciteSetBstMidEndSepPunct{\mcitedefaultmidpunct}
{\mcitedefaultendpunct}{\mcitedefaultseppunct}\relax
\EndOfBibitem
\bibitem[Invernizzi \latin{et~al.}(2020)Invernizzi, Piaggi, and
  Parrinello]{invernizzi2020unified}
Invernizzi,~M.; Piaggi,~P.~M.; Parrinello,~M. Unified approach to enhanced
  sampling. \emph{Phys. Rev. X} \textbf{2020}, \emph{10}, 041034\relax
\mciteBstWouldAddEndPuncttrue
\mciteSetBstMidEndSepPunct{\mcitedefaultmidpunct}
{\mcitedefaultendpunct}{\mcitedefaultseppunct}\relax
\EndOfBibitem
\bibitem[Invernizzi and Parrinello(2020)Invernizzi, and
  Parrinello]{invernizzi2020rethinking}
Invernizzi,~M.; Parrinello,~M. Rethinking metadynamics: from bias potentials to
  probability distributions. \emph{J. Phys. Chem. Lett.} \textbf{2020},
  \emph{11}, 2731--2736\relax
\mciteBstWouldAddEndPuncttrue
\mciteSetBstMidEndSepPunct{\mcitedefaultmidpunct}
{\mcitedefaultendpunct}{\mcitedefaultseppunct}\relax
\EndOfBibitem
\bibitem[Barducci \latin{et~al.}(2008)Barducci, Bussi, and
  Parrinello]{barducci2008well}
Barducci,~A.; Bussi,~G.; Parrinello,~M. Well-tempered metadynamics: a smoothly
  converging and tunable free-energy method. \emph{Phys. Rev. Lett.}
  \textbf{2008}, \emph{100}, 020603\relax
\mciteBstWouldAddEndPuncttrue
\mciteSetBstMidEndSepPunct{\mcitedefaultmidpunct}
{\mcitedefaultendpunct}{\mcitedefaultseppunct}\relax
\EndOfBibitem
\bibitem[Bussi and Laio(2020)Bussi, and Laio]{bussi2020using}
Bussi,~G.; Laio,~A. Using metadynamics to explore complex free-energy
  landscapes. \emph{Nat. Rev. Phys.} \textbf{2020}, \emph{2}, 200--212\relax
\mciteBstWouldAddEndPuncttrue
\mciteSetBstMidEndSepPunct{\mcitedefaultmidpunct}
{\mcitedefaultendpunct}{\mcitedefaultseppunct}\relax
\EndOfBibitem
\bibitem[Dama \latin{et~al.}(2014)Dama, Parrinello, and Voth]{dama2014well}
Dama,~J.~F.; Parrinello,~M.; Voth,~G.~A. Well-tempered metadynamics converges
  asymptotically. \emph{Phys. Rev. Lett.} \textbf{2014}, \emph{112},
  240602\relax
\mciteBstWouldAddEndPuncttrue
\mciteSetBstMidEndSepPunct{\mcitedefaultmidpunct}
{\mcitedefaultendpunct}{\mcitedefaultseppunct}\relax
\EndOfBibitem
\bibitem[Tiwary and Parrinello(2015)Tiwary, and Parrinello]{tiwary2015time}
Tiwary,~P.; Parrinello,~M. A time-independent free energy estimator for
  metadynamics. \emph{J. Phys. Chem. B} \textbf{2015}, \emph{119},
  736--742\relax
\mciteBstWouldAddEndPuncttrue
\mciteSetBstMidEndSepPunct{\mcitedefaultmidpunct}
{\mcitedefaultendpunct}{\mcitedefaultseppunct}\relax
\EndOfBibitem
\bibitem[Dama \latin{et~al.}(2015)Dama, Hocky, Sun, and
  Voth]{dama2015exploring}
Dama,~J.~F.; Hocky,~G.~M.; Sun,~R.; Voth,~G.~A. Exploring valleys without
  climbing every peak: more efficient and forgiving metabasin metadynamics via
  robust on-the-fly bias domain restriction. \emph{J. Chem. Theory Comput.}
  \textbf{2015}, \emph{11}, 5638--5650\relax
\mciteBstWouldAddEndPuncttrue
\mciteSetBstMidEndSepPunct{\mcitedefaultmidpunct}
{\mcitedefaultendpunct}{\mcitedefaultseppunct}\relax
\EndOfBibitem
\bibitem[Paszke \latin{et~al.}(2019)Paszke, Gross, Massa, Lerer, Bradbury,
  Chanan, Killeen, Lin, Gimelshein, Antiga, \latin{et~al.}
  others]{paszke2019pytorch}
Paszke,~A.; Gross,~S.; Massa,~F.; Lerer,~A.; Bradbury,~J.; Chanan,~G.;
  Killeen,~T.; Lin,~Z.; Gimelshein,~N.; Antiga,~L., \latin{et~al.}  Pytorch: An
  imperative style, high-performance deep learning library. \emph{Adv. Neur.
  Inf. Proc. Syst.} \textbf{2019}, \emph{32}\relax
\mciteBstWouldAddEndPuncttrue
\mciteSetBstMidEndSepPunct{\mcitedefaultmidpunct}
{\mcitedefaultendpunct}{\mcitedefaultseppunct}\relax
\EndOfBibitem
\bibitem[Pedregosa \latin{et~al.}(2011)Pedregosa, Varoquaux, Gramfort, Michel,
  Thirion, Grisel, Blondel, Prettenhofer, Weiss, Dubourg, Vanderplas, Passos,
  Cournapeau, Brucher, Perrot, and Duchesnay]{scikit-learn}
Pedregosa,~F. \latin{et~al.}  Scikit-learn: Machine Learning in {P}ython.
  \emph{J. Mach. Learn. Res.} \textbf{2011}, \emph{12}, 2825--2830\relax
\mciteBstWouldAddEndPuncttrue
\mciteSetBstMidEndSepPunct{\mcitedefaultmidpunct}
{\mcitedefaultendpunct}{\mcitedefaultseppunct}\relax
\EndOfBibitem
\bibitem[Piana \latin{et~al.}(2012)Piana, Lindorff-Larsen, and Shaw]{Piana2012}
Piana,~S.; Lindorff-Larsen,~K.; Shaw,~D.~E. {Protein folding kinetics and
  thermodynamics from atomistic simulation}. \emph{Proc. Natl. Acad. Sci. U. S.
  A.} \textbf{2012}, \emph{109}, 17845--17850\relax
\mciteBstWouldAddEndPuncttrue
\mciteSetBstMidEndSepPunct{\mcitedefaultmidpunct}
{\mcitedefaultendpunct}{\mcitedefaultseppunct}\relax
\EndOfBibitem
\bibitem[Piana \latin{et~al.}(2011)Piana, Lindorff-Larsen, and Shaw]{Piana2011}
Piana,~S.; Lindorff-Larsen,~K.; Shaw,~D.~E. {How robust are protein folding
  simulations with respect to force field parameterization?} \emph{Biophys. J.}
  \textbf{2011}, \emph{100}, L47--L49\relax
\mciteBstWouldAddEndPuncttrue
\mciteSetBstMidEndSepPunct{\mcitedefaultmidpunct}
{\mcitedefaultendpunct}{\mcitedefaultseppunct}\relax
\EndOfBibitem
\bibitem[Du \latin{et~al.}(1998)Du, Pande, Grosberg, Tanaka, and
  Shakhnovich]{du1998transition}
Du,~R.; Pande,~V.~S.; Grosberg,~A.~Y.; Tanaka,~T.; Shakhnovich,~E.~S. On the
  transition coordinate for protein folding. \emph{J. Chem. Phys.}
  \textbf{1998}, \emph{108}, 334--350\relax
\mciteBstWouldAddEndPuncttrue
\mciteSetBstMidEndSepPunct{\mcitedefaultmidpunct}
{\mcitedefaultendpunct}{\mcitedefaultseppunct}\relax
\EndOfBibitem
\bibitem[Bolhuis \latin{et~al.}(2002)Bolhuis, Chandler, Dellago, and
  Geissler]{bolhuis2002transition}
Bolhuis,~P.~G.; Chandler,~D.; Dellago,~C.; Geissler,~P.~L. Transition Path
  Sampling: Throwing Ropes. \emph{Annu. Rev. Phys. Chem} \textbf{2002},
  \emph{53}, 291--318\relax
\mciteBstWouldAddEndPuncttrue
\mciteSetBstMidEndSepPunct{\mcitedefaultmidpunct}
{\mcitedefaultendpunct}{\mcitedefaultseppunct}\relax
\EndOfBibitem
\bibitem[Karle and Balaram(1990)Karle, and Balaram]{karle1990structural}
Karle,~I.~L.; Balaram,~P. Structural characteristics of. alpha.-helical peptide
  molecules containing Aib residues. \emph{Biochem.} \textbf{1990}, \emph{29},
  6747--6756\relax
\mciteBstWouldAddEndPuncttrue
\mciteSetBstMidEndSepPunct{\mcitedefaultmidpunct}
{\mcitedefaultendpunct}{\mcitedefaultseppunct}\relax
\EndOfBibitem
\bibitem[Hartmann \latin{et~al.}(2020)Hartmann, Singh, Vanden-Eijnden, and
  Hocky]{Hartmann2020}
Hartmann,~M.~J.; Singh,~Y.; Vanden-Eijnden,~E.; Hocky,~G.~M. {Infinite switch
  simulated tempering in force (FISST)}. \emph{J. Chem. Phys.} \textbf{2020},
  \emph{152}, 244120\relax
\mciteBstWouldAddEndPuncttrue
\mciteSetBstMidEndSepPunct{\mcitedefaultmidpunct}
{\mcitedefaultendpunct}{\mcitedefaultseppunct}\relax
\EndOfBibitem
\bibitem[Buchenberg \latin{et~al.}(2015)Buchenberg, Schaudinnus, and
  Stock]{buchenberg2015hierarchical}
Buchenberg,~S.; Schaudinnus,~N.; Stock,~G. Hierarchical biomolecular dynamics:
  Picosecond hydrogen bonding regulates microsecond conformational transitions.
  \emph{J. Chem. Theory Comput.} \textbf{2015}, \emph{11}, 1330--1336\relax
\mciteBstWouldAddEndPuncttrue
\mciteSetBstMidEndSepPunct{\mcitedefaultmidpunct}
{\mcitedefaultendpunct}{\mcitedefaultseppunct}\relax
\EndOfBibitem
\bibitem[Biswas \latin{et~al.}(2018)Biswas, Lickert, and
  Stock]{biswas2018metadynamics}
Biswas,~M.; Lickert,~B.; Stock,~G. Metadynamics enhanced Markov modeling of
  protein dynamics. \emph{J. Phys. Chem. B} \textbf{2018}, \emph{122},
  5508--5514\relax
\mciteBstWouldAddEndPuncttrue
\mciteSetBstMidEndSepPunct{\mcitedefaultmidpunct}
{\mcitedefaultendpunct}{\mcitedefaultseppunct}\relax
\EndOfBibitem
\bibitem[Abraham \latin{et~al.}(2015)Abraham, Murtola, Schulz, P{\'a}ll, Smith,
  Hess, and Lindahl]{abraham2015gromacs}
Abraham,~M.~J.; Murtola,~T.; Schulz,~R.; P{\'a}ll,~S.; Smith,~J.~C.; Hess,~B.;
  Lindahl,~E. GROMACS: High performance molecular simulations through
  multi-level parallelism from laptops to supercomputers. \emph{SoftwareX}
  \textbf{2015}, \emph{1}, 19--25\relax
\mciteBstWouldAddEndPuncttrue
\mciteSetBstMidEndSepPunct{\mcitedefaultmidpunct}
{\mcitedefaultendpunct}{\mcitedefaultseppunct}\relax
\EndOfBibitem
\bibitem[Ferrarotti \latin{et~al.}(2015)Ferrarotti, Bottaro, P{\'e}rez-Villa,
  and Bussi]{ferrarotti2015accurate}
Ferrarotti,~M.~J.; Bottaro,~S.; P{\'e}rez-Villa,~A.; Bussi,~G. Accurate
  multiple time step in biased molecular simulations. \emph{J. Chem. Theory
  Comput.} \textbf{2015}, \emph{11}, 139--146\relax
\mciteBstWouldAddEndPuncttrue
\mciteSetBstMidEndSepPunct{\mcitedefaultmidpunct}
{\mcitedefaultendpunct}{\mcitedefaultseppunct}\relax
\EndOfBibitem
\bibitem[Huang \latin{et~al.}(2017)Huang, Rauscher, Nawrocki, Ran, Feig,
  De~Groot, Grubm{\"u}ller, and MacKerell]{huang2017charmm36m}
Huang,~J.; Rauscher,~S.; Nawrocki,~G.; Ran,~T.; Feig,~M.; De~Groot,~B.~L.;
  Grubm{\"u}ller,~H.; MacKerell,~A.~D. CHARMM36m: an improved force field for
  folded and intrinsically disordered proteins. \emph{Nat. Methods}
  \textbf{2017}, \emph{14}, 71--73\relax
\mciteBstWouldAddEndPuncttrue
\mciteSetBstMidEndSepPunct{\mcitedefaultmidpunct}
{\mcitedefaultendpunct}{\mcitedefaultseppunct}\relax
\EndOfBibitem
\end{mcitethebibliography}

\setcounter{figure}{0} 
\setcounter{equation}{0}    
\setcounter{section}{0}    

\renewcommand{\thesection}{S\arabic{section}}
\renewcommand{\thefigure}{S\arabic{figure}}
\renewcommand{\thetable}{S\arabic{table}}
\renewcommand{\theequation}{S\arabic{equation}}

\onecolumn
\begin{center}
    \LARGE Supporting Information
\end{center}

\begin{figure}
    \centering
    \includegraphics[width=5in]{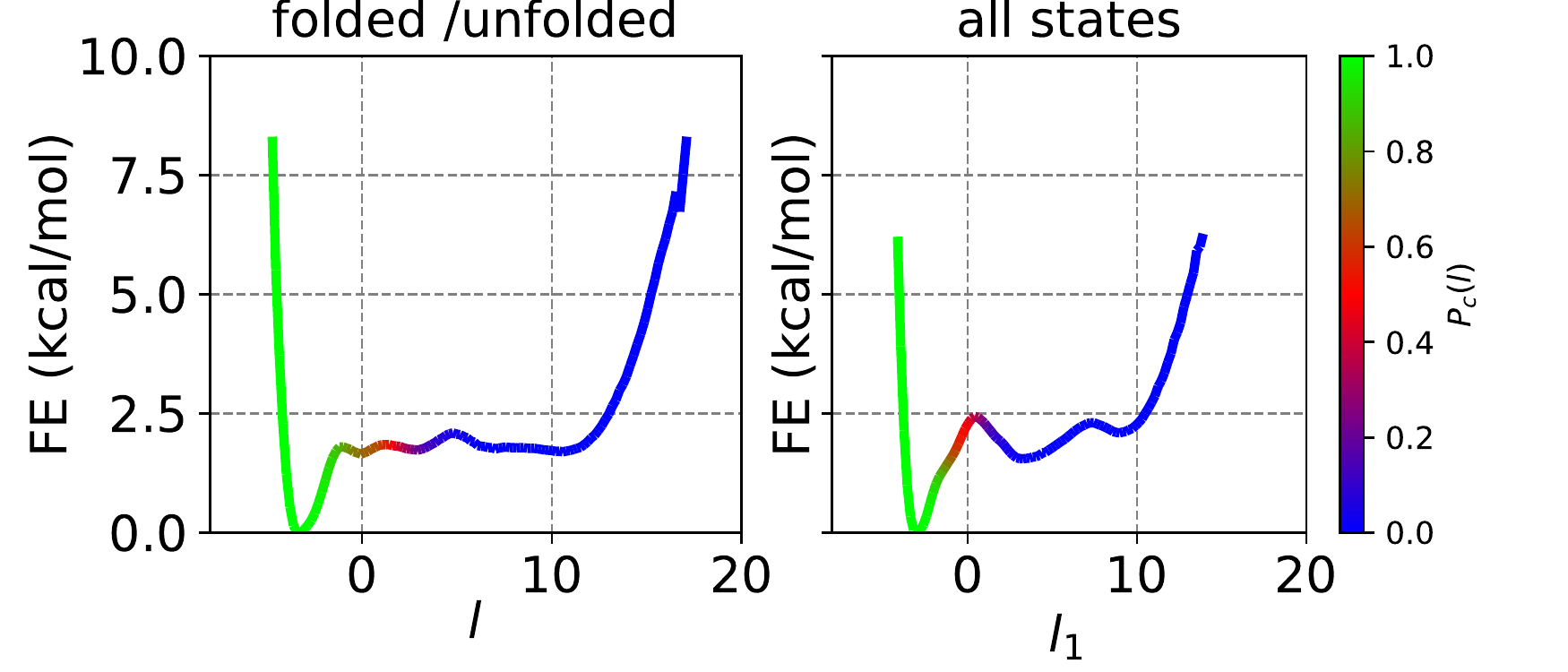}
    \caption{ FES obtained from unbiased MD colored by committor probability. The left shows the FES computed along $l$, the LDA coordinate from states 0 and 4 in our GMM model, and the right shows the FES computed along $l_1$, the first LDA coordinate from a model trained on all 6 states. }
    \label{fig:compare_hp35_2state_6state}
\end{figure}

\begin{figure}[ht]
    \centering
    \includegraphics[width=5in]{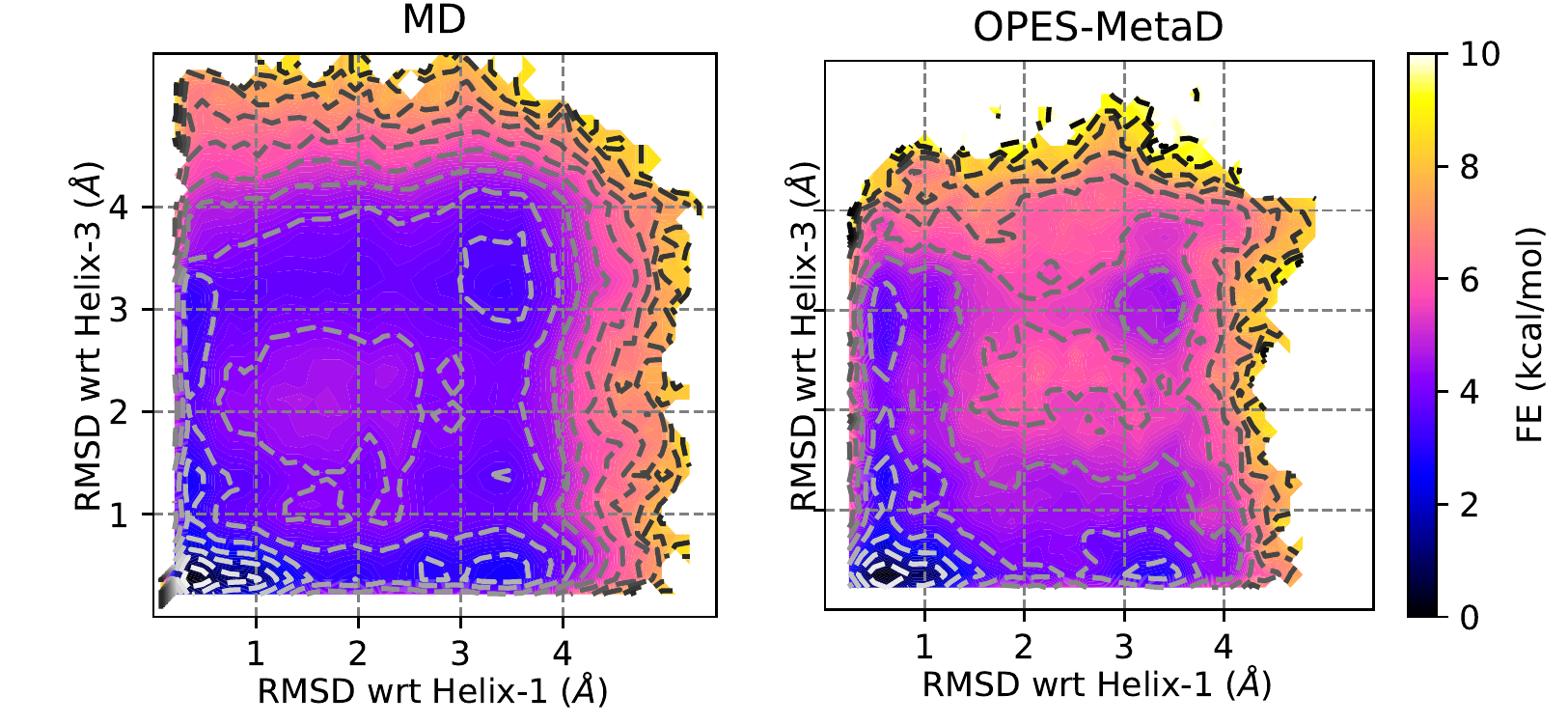}
    \caption{FES from unbiased MD simulation and OPES-MetaD reweighted along coordinates measuring RMSD of the terminal two helices to a reference folded structure. }
    \label{fig:hp35_fe_rmsd}
\end{figure}

\begin{figure}
    \centering
    \includegraphics[width=7in]{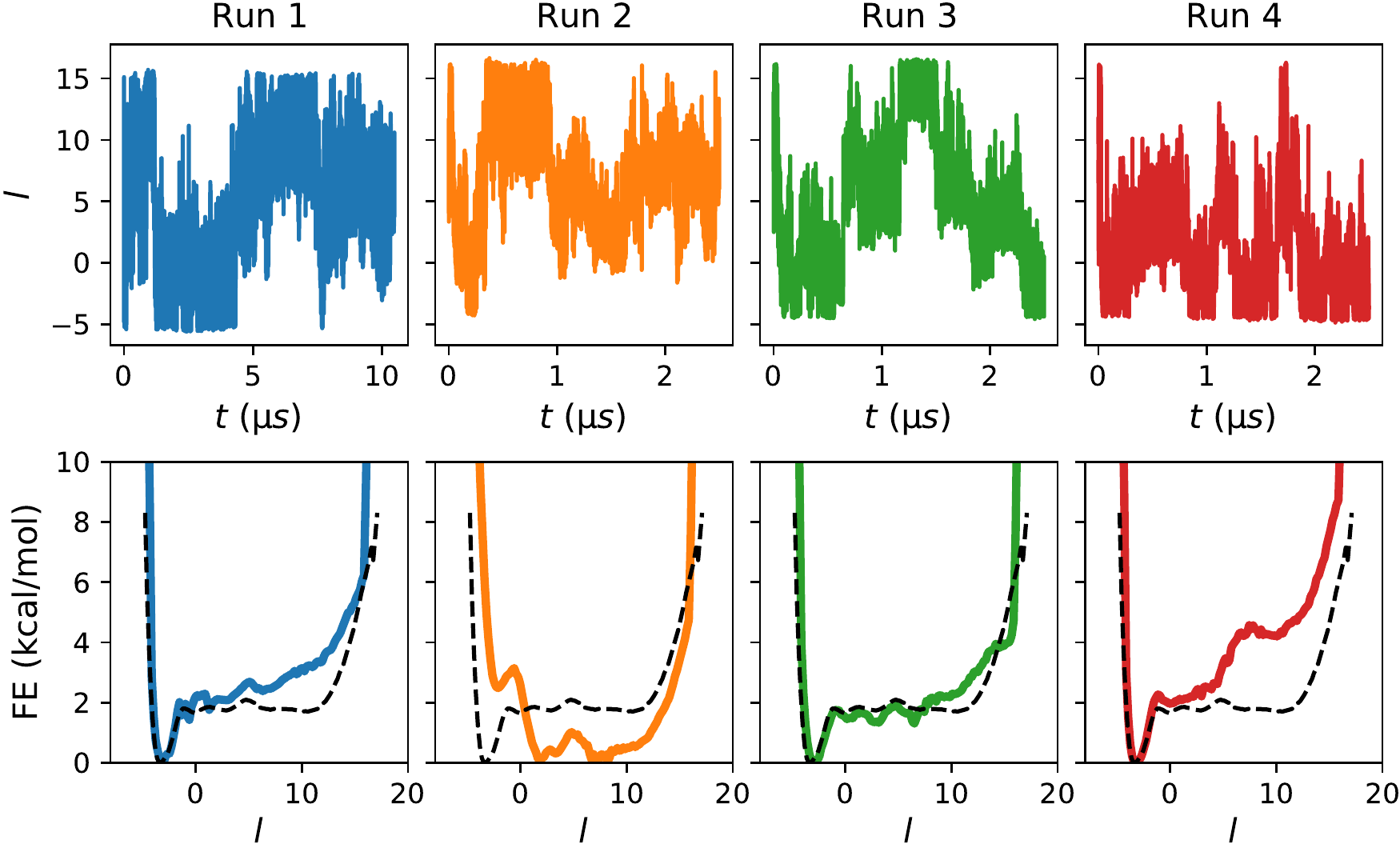}
    \caption{(top) OPES-MetaD replicate trajectories for HP35. 
    Run 1 is the 10$\mathrm{mu}s$ simulation studied in the main text.
    The other three runs are 2.5$\mathrm{mu}s$ runs starting from other points obtained in the original trajectory, with separate equilibrium performed. Each of these three simulations has one to two transitions.
    (bottom) Comparison of FES obtained from OPES-MetaD for HP35, with a dashed line showing the FES obtained from unbiased MD. Run 3 producing a perfect FES by chance. }
    \label{fig:hp35_compare_four_runs}
\end{figure}

\end{document}